\documentclass[aps, twocolumn, showpacs, amsmath]{revtex4}
\usepackage{dcolumn}
\usepackage{bm}
\usepackage{graphicx}
\usepackage{color}
\usepackage{subfigure}
\usepackage{hyperref}
\usepackage{latexsym}
\usepackage{amsthm}
\usepackage{amssymb}

\DeclareGraphicsExtensions{.jpg,.pdf, .mps, .png, .eps, .ps, .EPS,.gif}

\begin{document}
\def\be{\begin{equation}}
\def\ee{\end{equation}}

\def\bc{\begin{center}}
\def\ec{\end{center}}
\def\bea{\begin{eqnarray}}
\def\eea{\end{eqnarray}}
\newcommand{\avg}[1]{\langle{#1}\rangle}
\newcommand{\Avg}[1]{\left\langle{#1}\right\rangle}

\def\ie{\textit{i.e.}}
\def\etal{\textit{et al.}}
\def\m{\vec{m}}
\def\G{\mathcal{G}}

\title{Multiple percolation transitions in a configuration model of  network of networks }
\author{  Ginestra Bianconi}

\affiliation{
School of Mathematical Sciences, Queen Mary University of London, London, E1 4NS, United Kingdom}
\author{ Sergey N. Dorogovtsev}
\affiliation{Departamento de F\'{\i}sica da Universidade de Aveiro, I3N, 3810-193, Aveiro, Portugal\\
A. F. Ioffe Physico-Technical Institute, 194021 St.~Petersburg, Russia}

\begin{abstract}
Recently 
much attention has been  
paid to the study of the robustness of interdependent 
and multiplex networks and, in particular, networks of networks. The robustness of interdependent networks can be evaluated by 
the  size of  
a mutually connected component  
when a fraction 
of nodes have been removed from 
these networks. 
Here we characterize the emergence of the mutually connected component in a network of networks in which every node of a network (layer) $\alpha$ is connected   with  $q_{\alpha}$ its 
randomly chosen replicas  in some other networks and is interdependent of these nodes with probability $r$. We find that when the superdegrees $q_{\alpha}$ of 
different layers 
in the network of networks are distributed  
heterogeneously, multiple percolation phase transition 
can occur. 
We show that, depending on the value of $r$, these transition 
are continuous or discontinuous.

\end{abstract}

\pacs{89.75.Fb, 64.60.aq, 05.70.Fh, 64.60.ah}

\maketitle


\section{Introduction}

The complexity of a large variety of  systems, from  infrastructures to the cell, is rooted in 
a network of 
interactions between their constituents \cite{RMP,Newman_rev,Boccaletti2006}. Quantifying the robustness of complex networks is one of the main challenges of network of networks with implications in fields as different as biology or policy making and risk assessment.

In the last fifteen years it has been shown \cite{crit,Dynamics} that the structure of a single network is strictly related to its robustness.
But only recently \cite{Havlin1,Arenas}, attention has been drawn toward a previously neglected aspects of complex systems, 
namely the  interactions between several complex networks.

Rarely single networks are isolated, while  it is usually  the case that several networks are interacting and  interdependent on each other. For example, in infrastructures, the banking systems are interdependent  with the Internet and the electric power-grid,  public transport, such as subway, is dependent on the power-grid, which relies on its turn on  the water supply system to cool the power-plants, etc. In the cell the situation is much similar: all cellular networks, such as the metabolic networks, the protein-protein interaction networks, the signaling networks, and the gene transcription networks are all dependent on each other, and the cell is only alive if 
all  these networks are functional.  
These are  examples  of network of networks, i.e., networks formed by several 
interdependent networks.

A special class of network of networks are multiplex networks \cite{Arenas,Mucha,Thurner,Boccaletti}, which are multilayer structures 
in which each layer is formed by the same set of nodes $N$ interconnected by different kinds of links for different layers. 
In other words, these are graphs with all nodes of one kind and with links of different colors. 
Multiplex networks are 
attracting 
great interest as they represent a large variety of systems such as social networks where people can be linked by different types of relationships (friendship, family tie, collaboration, citations, etc.)  or, for example, in transportation networks, where different places can be linked by different types of  transportation  
(train, flight connections, flight connection of different airline companies, etc.).
Multiplex network datasets \cite{Mucha,Thurner,Boccaletti} are starting to be analysed, several modelling framework for these networks have been proposed \cite{PRL,Growth,PRE} and the characterization of a large variety of dynamical processes is getting a momentum \cite{Diffusion,Boguna,dedomenico,Cooperation,Radicchi:ra13,Dickison:dhs2012,Sole-Ribalta:sdkdga2013}. 
At this point we emphasize the principal difference between the interdependent and so-called interconnected networks. This difference is not about structural organization of connections between these networks but rather about the function of these interlinks. Interlinks connecting different interdependent networks (interdependencies) show the pairs of nodes that cannot exist without each other. In the present paper we consider variations of this kind of complex networks. On the other hand, in the interconnected networks \cite{Leicht:ls2009,Dickison:dhs2012,Radicchi:ra13,Sole-Ribalta:sdkdga2013}, the interlinks play the same role as links in single networks, enabling one to consider, e.g., various percolation problems, disease spreading, etc.

A major progress in understanding the robustness of multilayer interdependent networks has been made in a series of seminal papers \cite{Havlin1,parshani,HavlinPRL,Son}, where it has been proposed that a 
natural measure for evaluating the robustness of these structures  to random failure is  the size of 
a mutually connected giant component.
The mutually connected giant component is the component that remains after breakdowns propagate back and forth between 
different interdependent networks (layers) generating a cascade of failure events.
A node is in the mutually connected component of a multilayer network if all the nodes on which it depends are also in the mutually connected network and if at least one neighbor node in its own network (layer) belongs to the mutually connected component \cite{Son,Doro,Kabashima,DeDominico:dsckmpga2013}. 
Clearly, the giant mutually connected component naturally generalizes the giant connected component (percolation cluster) in a single network. 
The robustness properties of multiplex networks have been right now well understood \cite{Havlin1, parshani, HavlinPRL,Son,Doro,Kabashima}, including 
effects of degree correlations, 
the 
overlap of the links or antagonistic effects in this novel type of percolation problem \cite{Goh,PREoverlap,JSTAT}.
As the fraction of $1-p$ of removed nodes---``igniters''---increases, multiplex networks are affected by cascading failures, until they reach a point for $p=p_c$ where the network abruptly collapses, and the size of the mutually connected component shows a discontinuous transition \cite{Havlin1,parshani,HavlinPRL,Son,Doro}.
In this case if a small fraction $1-r$ of nodes in the multiplex are not interdependent, then the transition can  
change from discontinuous to continuous \cite{HavlinPRL}. 
Although the issue of interest in the present article is the giant mutually connected component, other special giant components can be introduced for these networks. Here we mention only the so-called giant viable cluster \cite{Son,Doro} between each two nodes of which, there is a complete set of interconnecting paths running through every layer. It is easy to see that the viable cluster is a subgraph of the mutual component. 

\begin{figure}[t]
\begin{center}
\includegraphics[scale=.303]{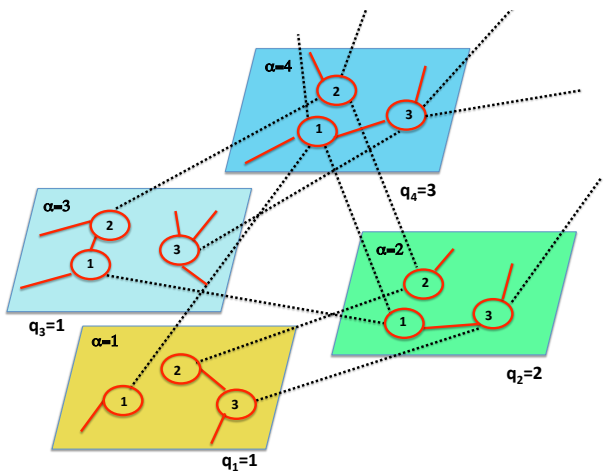}
\end{center}
\caption{
(Color online) Schematic view of the configuration model of a network of
networks in the case of $r=1$. Interdependencies (interlinks between nodes from different levels) are shown by the black dashed lines. Intralinks between nodes within layers are
shown as solid red lines. In each individual layer (label $\alpha=1,2,3,4,\ldots,M$), all nodes ($i=1,2,3,\ldots,N$) have the same number of interlinks (superdegree $q_\alpha$). Interlinks connect only nodes with the same label $i$ in different layers, forming $N$ ``local supernetworks'' $i=1,\ldots,N$. Each of these local supernetworks is an uncorrelated random graph with a given superdegree sequence $\{q_\alpha\}$, $\alpha=1,\ldots,M$, defined as the standard configuration model (uniformly random interconnections for a given sequence of superdegrees). 
}
\label{f01}
\end{figure}

Several works have 
considered 
afterwards 
a more general (than multiplex) case of a network of networks 
in which the layers are formed by the same number of nodes, 
where each node has a replica node in each other network and 
might depend only  on its replica nodes.
Despite the major interest on the topic \cite{Havlin3,Gao:gbhs2012,Gao:gbhs2011,Dong:dtdxzs2013,Gao:gbsxh2013,Dong:dgdtsh2013}, only recently 
the following key circumstance become clear \cite{BD}.  
When the interdependencies between the nodes are such that if a network depends on another network, all its nodes depend on the replica nodes of the other network, 
it turns out that the mutually connected component coincides with that in the corresponding fully connected network of networks, which is actually a particular case of a multiplex network.

Here we show, nevertheless, that the situation changes if the interdependence links 
are distributed between layers more randomly, i.e., if we remove the above constraint that all nodes in each layer are dependent on their replicas in the same set of other layers. 
We consider the situation in which 
a superdegree
$q_{\alpha}$ is assigned to each layer $\alpha$, 
so that each node of the network depends only on other $q_{\alpha}$ replica nodes, but now these replica nodes 
are chosen randomly and independently from different layers, see Fig.~\ref{f01}. We call this 
construction the configuration model of a network of networks. 
The reason for this term is that 
for each set of nodes with given $i$, consisting of nodes $(i,\alpha)$ in all the layers $\alpha=1,\ldots,M$, our definition provides the standard configuration model of a random network with a given sequence of superdegrees $q_\alpha$, $\alpha=1,\ldots,M$, (``local supernetwork'' $i$). 
Let us compare this construction to the model of Ref.~\cite{BD}. While in Ref.~\cite{BD}, all $N$ ``local supernetworks'' coincided, in the configuration model of network of networks defined in the present work, the ``local supernetworks'' differ from each other; only their superdegrees $q_\alpha$ coincide in the particular case of $r=1$. Depending on the superdegree sequence, ``local supernetworks'' can contain a set of finite connected components and a giant one.   
We consider specifically the case in which all the layers have the same internal degree distribution $P(k)$  and where the superdegree distribution is $P(q)$. 
We derive equations for the order parameter of this problem. Solving them numerically and analytically we show for this network of networks 
that if  $r=1$, the layers with increasing superdegree $q$ have a percolation transition at different values of $p$, where $p$ is the fraction of not damaged nodes in the network, see Fig.~\ref{fig:fig1a}. 
In other words, as $p$ increases, in layers with higher and higher $q$, giant clusters of the mutual component emerge progressively, or, one can also say, the mutual component expands to layers with higher $q$, see Fig.~\ref{f2}. 
All these transition are abrupt  
but also they are characterized by a singularity of the order parameter. Finally we show that for $r<1$, 
i.e. when interlinks between layers are removed with probability $1-r$, 
each of these transitions involving the layers with $q_{\alpha}=q$ can become continuous for $r<r_q$. 
In the case of $r<1$, nodes in the same layer have different number of interlinks to other layers, which essentially generalizes the problem.


\section{Percolation in Network of Networks: The message passing approach}

 A network of networks is formed by $M$ networks (layers), $\alpha=1,2\ldots, M$, each formed by  $N$  nodes, $i=1,2\ldots, N$, where $N$ is infinite. In fact, in this work we also set $M\to \infty$, which enables us to use an analytical technique developed for the standard configuration model. These calculations exploit the locally tree-like structure of infinite uncorrelated networks. A more realistic case of a finite $M$ is a challenging problem. 
Every node $(i,\alpha)$ 
is connected  
with a number of nodes $(j,\alpha)$ in 
the same 
layer 
and with a number of  
its ``replica nodes'' $(i,\beta)$ in other layers. 
For two layers, this framework was proposed in Ref.~\cite{Diffusion}. It was generalized to an arbitrary number of layers in Refs.~\cite{DeDominico:dsckmpga2013,Sole-Ribalta:sdkdga2013}. 
We consider the situation in which each 
layer $\alpha$ is interdependent on $q_{\alpha}$ other networks, so that each node $i$ in layer $\alpha$ has exactly $q_\alpha$ interlinks, which connect node $i$ only to its replica nodes in $q_\alpha$ of other layers. We call $q_{\alpha}$ the superdegree of
the nodes of 
layer $\alpha$. 
We stress that the superdegree 
is associated with the nodes of a layer, and each node $(i,\alpha)$ in a layer has the same superdegree $q_\alpha$. 
Interlinks connecting replica nodes within different sets, say, interlink $(i\alpha-i\beta)$ and interlink $(j\gamma-j\delta)$ are assumed to be independent (uncorrelated). 
For example, if a  node $(i,\alpha)$ of layer $\alpha$ is interdependent on node $(i,\beta)$ of layer $\beta$, then 
although another node $(j,\alpha)$ of layer $\alpha$ 
may in principle 
occur interdependent on node $(j,\beta)$, in general, it depends on $q_{\alpha}$ replica nodes $(j,\gamma)$ sitting in any $q_{\alpha}$ layers. 
Following Refs.~\cite{Diffusion,DeDominico:dsckmpga2013}, we define the network of networks with a super-adjacency matrix of elements $a_{i\alpha,j\beta}=1$ if there is a link between node $(i,\alpha)$ and node $(j,\beta)$ and zero otherwise. In these networks we have always $a_{i\alpha,j\beta}=0$ if both $i\neq j$ and $\alpha\neq \beta$. 
For each node $i$ and all its replicas, we introduce a ``local supernetwork'', whose nodes are the layers and the links are interdependencies within this set of replicas. 
This network was discussed in Refs.~\cite{Sole-Ribalta:sdkdga2013,dedomenico}. 
This local supernetwork is determined by the adjacency matrix $A^i_{\alpha,\beta}=((a_{i\alpha,i\beta}))$ parametrized by the node $i$. This network may consist of a number of connected components. Connected components from different local supernetworks are connected with each other through links within individual layers. In this work we explore this complicated system of interconnected components, which is necessary to describe the emergence of the giant mutually connected component and its expansion (percolation) over different layers.  

We define the mutually connected component as the following.
Each node $(i,\alpha)$ is in the mutually connected component if it has at least one neighbor $(j,\alpha)$ which belongs to the mutually connected component and if all the linked nodes $(i,\beta)$ in the interdependent networks are also in the mutually connected component. 
In these problems, the giant (i.e., containing a finite fraction of nodes) mutually connected component is single. 
It immediately follows from this definition that, remarkably, within each connected component of any local supernetwork, all its replica nodes either together belong to the mutually connected component or not. In our considerations, we will essentially exploit this strong consequence.

Given a network of networks it is easy to construct a message passing algorithm \cite{Son,Kabashima,Mezard,Weigt} determining  if node $(i,\alpha)$ is in the mutually connected component. Let us denote by $\sigma_{i\alpha\to j\alpha}=1,0$ the  message 
within a layer, from node $(i,\alpha)$ to node $(j,\alpha)$ and indicating ($\sigma_{i\alpha\to j\alpha}=1$) if node $(i,\alpha)$ is in the mutually connected component when we remove the link $(i,j)$ in network $\alpha$. Furthermore, let us denote by $S'_{i\alpha\to i\beta}=0,1$ the  message between the ``replicas'' $(i,\alpha)$ and $(i,\beta)$ of node $i$ in layers $\alpha$ and $\beta$. The message $S'_{i\alpha\to i\beta}=1$ indicates if the node $(i,\alpha)$ is in the mutually connected component when we remove the link between node $(i,\alpha)$ and node $(i,\beta)$.
 In addition to that we  assume that the node $(i,\alpha)$ can be damaged and permanently removed from the network. 
This removal will launch an avalanche of failures (removals of nodes) spreading over the layers. Note that, of course, the node removal retains its interdependence links.
We indicate with  $s_{i\alpha}=0$ a node that is damaged, otherwise we have $s_{i\alpha}=1$.
The message passing equations for these messages are given by 
\bea
\sigma_{i\alpha\to j\alpha}&=&s_{i\alpha}\prod_{\beta\in {\cal N}_i(\alpha)}S'_{i\beta\to i\alpha}
\nonumber \\[5pt]
&&\times\left[1-\prod_{\ell\in N_{\alpha}(i)\setminus j}(1-\sigma_{\ell\alpha\to i\alpha})\right]
,
\nonumber \\[5pt]
S'_{i\alpha\to i\beta}&=&s_{i\alpha}\prod_{\gamma\in {\cal N}_i(\alpha)\setminus \beta}S'_{i\gamma\to i\alpha}
\nonumber \\[5pt]
&&\times \left[1-\prod_{\ell\in N_{\alpha}(i)}(1-\sigma_{\ell\alpha\to i\alpha})\right]
,
\label{mp1}
\eea
where $N_{\alpha}(i)$ indicates the set of  nodes $(\ell,\alpha)$ which are neighbors of node $i$ in network $\alpha$, and ${\cal N}_i(\alpha)$ indicates the layers  $\beta$ such that the nodes $(i,\beta)$ are interdependent on the node $(i,\alpha)$.
Finally $S_{i\alpha}$ indicates if   a node $(i,\alpha)$ is in the mutually interdependent network ($S_{i\alpha}=1,0$) and this indicator function can be expressed in terms of the messages as 
\bea
S_{i\alpha}&=&s_{i\alpha}\prod_{\beta\in {\cal N}_i(\alpha)}S'_{i\beta\to i\alpha}
\nonumber \\[5pt]
&&\times \left[1-\prod_{\ell\in N_{\alpha}(i)}(1-\sigma_{\ell\alpha \to i\alpha})\right]
.
\label{S}
\eea 
The solution of the message passing equations is given by the following closed expression,
\bea
\hspace*{-2mm}\sigma_{i\alpha\to j\alpha}&=&
\!\!\!\prod_{\gamma\in {\cal C}({i,\alpha})\setminus \alpha}\!\left\{s_{i\gamma}\left[1-\prod_{\ell\in N_{\gamma}(i)}(1-\sigma_{\ell\gamma\to i\gamma})\right]\right\}
\nonumber \\[5pt]
\hspace*{-10mm}&&\times s_{i\alpha}\left[1-\prod_{\ell\in N_{\alpha}(i)\setminus j}(1-\sigma_{\ell\alpha\to i\alpha})\right],
\label{mesf}
\eea
where ${\cal C}(i,\alpha)$ is the connected cluster of the ``local supernetwork'' of node $i$, i.e., the network between layers determined by the adjacency matrix $A^i_{\alpha,\beta}=((a_{i\alpha,i\beta}))$ parametrized by the node $i$. 
Finally $S_{i\alpha}$ is given by 
\bea
S_{i\alpha}&=&s_{i\alpha}\prod_{\beta\in {\cal C}(i,\alpha)}\left[1-\prod_{\ell\in N_{\beta}(i)}(1-\sigma_{\ell\beta \to i\beta})\right].
\label{S2}
\eea 
For detailed derivation and explanation of this solution see our work \cite{BD} and Appendix.


\section{Percolation in the Configuration Model of  Network of Networks}
\label{s3}

 We assume here that each network 
(layer) $\alpha$ is generated from a configuration model with  the same degree distribution $P_{\alpha}(k)=P(k)$, and that each node $(i,\alpha)$ is connected to $q_{\alpha}$ other ``replica'' nodes $(i,\beta)$ chosen uniformly randomly.  
Moreover we assume that the degree sequence in  each layer is $\{k_i^{\alpha}\}$ and that the degrees of the replicas of node $i$ are uncorrelated.
This implies that we are considering a network of networks ensemble, such that every network of networks with a super-adjacency matrix ${\bf a}$ has a probability $P({\bf a})$ given by 
\bea
\hspace*{-20mm}P({\bf a})&=&\frac{1}{Z}\prod_{\alpha=1}^M\prod_{i=1}^N\left\{\delta\left(k^{\alpha}_i,\sum_{j=1}^Na_{i\alpha,j\alpha}\right)\delta\left(q_{\alpha},\sum_{\beta=1}^Ma_{i\alpha,i\beta}\right)\right.
\nonumber \\[5pt]
&&\left.\times\left[\prod_{\beta\neq \alpha}\prod_{j\neq i}\delta\left(a_{i,\alpha,j\beta},0\right)\right]\right\}, 
\label{Pa}
\eea
where $\delta(a,b)$ indicates the Kronecker delta and $Z$ is a normalization constant.
Moreover we assume that  
nodes  $(i,\alpha)$ are removed with probability $1-p$, i.e., we consider  the following expression for the probability $P(\{s_{i\alpha}\})$ of the variables $s_{i\alpha}$   
\bea
P(\{s_{i\alpha}\})=\prod_{\alpha=1}^M\prod_{i=1}^N p^{s_{i\alpha}}(1-p)^{1-s_{i\alpha}}.
\label{Ps}
\eea
In order to quantify the expected  size of the mutually connected component in this ensemble, we can average the messages over this ensemble of the network of networks.
The message passage equations for this problem are given by Eqs.~(\ref{mesf}) .
Therefore the equations for the average message 
within a layer are given in terms of the parameter $p=\Avg{s_{i\alpha}}$ and the generating functions $G_0^{k}(z),G_1^{k}(z),G_0^q(z), G_1^q(z)$ given by 
\bea
G_0^k(z)=\sum_k P(k) z^k
,\  & 
G_1^k(z)=\displaystyle\sum_k \displaystyle\frac{kP(k)}{\Avg{k}}z^{k-1}
,
\nonumber 
\\[5pt]
G_0^q(z)=\sum_q P(q) z^q
,\  & 
G_1^q(z)=\displaystyle\sum_q \displaystyle\frac{qP(q)}{\Avg{q}}z^{q-1}.
\eea
In particular, if we indicate by $\sigma_q$ the average messages 
within a layer $\alpha$ of degree $q_{\alpha}=q$ we obtain 
\bea
\sigma_{q}&=&p\sum_{s}P(s|q)\left[p\sum_{q'} \frac{q' P(q')}{\Avg{q}}
[1-G_0(1-\sigma_{q'})]\right]^{s-1}
\nonumber \\[5pt]
&&\times[1-G_1(1-\sigma_{q})],
\label{sq}
\eea
where $P(s|q)$ indicates the probability that a node $i$ in layer $\alpha$ with $q_{\alpha}=q$ is in a connected component ${\cal C}(i,\alpha)$ of the local supernetwork
of cardinality (number of nodes) $|{\cal C}(i,\alpha)|=s$.
Similarly, the probability that a node $i$ in a layer $\alpha$ with superdegree $q_{\alpha}=q$ is in the mutually connected component $S_q=\Avg{S_{i\alpha}}$ is given by 
\bea
S_q&=&p\sum_{s}P(s|q)\left[p\sum_{q'} \frac{q' P(q')}{\Avg{q}}
[1-G_0(1-\sigma_{q'})]\right]^{s-1}
\nonumber \\[5pt]
&&\times[1-G_0(1-\sigma_{q})].
\label{Sq}
\eea
Equations~(\ref{sq}) and (\ref{Sq}) are valid for any network of networks ensemble described by Eqs.~(\ref{Pa})--(\ref{Ps}). In the following we study in particular 
the limiting case 
in which the number of layers $M\to \infty$, and 
the local supernetwork is sparse. 
In order to find a solution for Eqs.~(\ref{sq}), we define $B$ as 
\bea
B=
p
\sum_{q'}  \frac{q' P(q')}{\Avg{q}}\,
[1-G_0(1-\sigma_{q'})]
.
\eea
Inserting this expression in Eq.~(\ref{sq}) we get 
\bea
\sigma_{q}&=&p
\sum_{s}P(s|q)B^{s-1}
\,\,\,
[1-G_1(1-\sigma_{q})].
\label{sq2}
\eea
From the definition of $B$ we see that $B\leq 1$ and that $B=1$ only if both $p=1$ and $\sigma_{q}=1\ \forall q$. 
This implies that in all the cases in which the layers are not all formed by a single giant component  or in which $p<1$, we have  $B<1$. Moreover if $B<1$,  we can neglect in Eq.~(\ref{sq}) the contribution coming from the giant component of the local supernetworks in the large $M$ limit.  Therefore in Eq.~(\ref{sq2}) we can 
replace $P(s|q)$ with the probability $P_f(s|q)$ that a node of degree $q$  belongs to a finite component of size $s$ in the supernetwork. 
Note that in our model, the statistics of all local supernetworks coincide. 
Let us consider the quantity
$P(s)=\sum_{q}[qP(q)/\avg{q}]P_f(s|q)$. 
Since $P(s)$ only depends on the distribution of finite components, we have 
\bea
\!\!\!\!
P(s)=\sum_{q}\frac{qP(q)}{\Avg{q}}\!\!\sum_{s_1,s_2,\ldots s_q}\prod_{\ell=1}^qP(s_{\ell})\delta\!\left(\sum_{\ell=1}^qs_{\ell},s-1\!\!\right)\!.
\eea
Therefore $P(s)$
has 
the generating function $H(z)=\sum_{s}P(s)z^s$ that satisfies the equation
\bea
H(z)=z G^q_1(H(z)).
\eea
Moreover,
since 
\bea
P_f(s|q)=\sum_{s_1,s_2,\ldots s_q}\prod_{\ell=1}^qP(s_{\ell})\,\delta\!\left(\sum_{\ell=1}^qs_{\ell},s-1\!\right),
\eea
we find 
\bea
\sum_{s}P_f(s|q)B^{s-1}=[H(B)]^{q}.
\eea
Therefore Eqs.~(\ref{sq}) become
\bea
\sigma_q&=&p\,\, [H(B)]^q [1-G_1^k(1-\sigma_q)]
, 
\nonumber
\\[5pt]
H(z)&=&zG^q_1(H(z))
, 
\nonumber
\\[5pt]
B&=&
p\sum_{q'}\frac{q' P(q')}{\Avg{q}}
\,[1-G_0^k(1-\sigma_{q'})]
.
\label{e150}
\eea
Putting $H(B)=\Sigma$, 
we can express Eqs.~(\ref{sq}) and Eqs.~(\ref{Sq}) in the following simplified way,
\bea
\sigma_q&=&p\,\,(\Sigma)^q[1-G_1^k(1-\sigma_q)] 
,
\nonumber
\\[5pt]
S_q&=&p\,\,(\Sigma)^q[1-G_0^k(1-\sigma_q)] 
,
\nonumber
\\[5pt]
\Sigma &=&\left[p\sum_{q'}\frac{q' P(q')}{\Avg{q}}
[1-G_0^k(1-\sigma_{q'})]\right]
\nonumber 
\\[5pt]
&&\times \sum_{q'}\frac{q' P(q')}{\Avg{q}}(\Sigma)^{q'-1}
.
\label{e160}
\eea
Therefore, given a configuration model of a network of networks, the parameter $\Sigma$ determines both $\sigma_q$ and $S_q$ for any value of the superdegree $q$. For this reason we can call $\Sigma$ 
the  order parameter for the entire network of networks. 
Let us consider for simplicity the case in which each layer is formed by a Poisson network with average degree $\avg{k}=c$. In this situation the previous equations become
\bea
\sigma_q
&=&
S_q=
p\,\,
(\Sigma)^q(1-e^{-c\sigma_q}) 
,
\nonumber
\\[5pt]
&&\hspace{-49pt}
\Sigma=\!
\left[p\sum_{q'}\frac{q' P(q')}{\Avg{q}}
(1-e^{-c\sigma_{q'}})\right]
\!\!\sum_{q'}\frac{q' P(q')}{\Avg{q}}(\Sigma)^{q'-1}
.
\label{ps}
\eea

In particular, if
the local supernetwork is regular, i.e., $P(q)=\delta(q,m)$ we have $\sigma=\sigma_m=\sqrt{\Sigma}$  satisfying, 
\bea
\sigma=p\sigma^{q/2}(1-e^{-c\sigma})
.
\eea
This special random regular network of interdependent Poisson networks was recently considered in Ref.~\cite{Gao:gbsxh2013}. 
In the  
general case of an arbitrary 
$P(q)$ distribution the problem defined in Eqs.~(\ref{ps}) continues to have  a single order parameter given by $\Sigma$. In fact the  first equation in Eqs.~(\ref{ps}) has a solution expressed in terms of the principal value of the Lambert function $W(x)$, which is given by 
\bea
\sigma_q=\frac{1}{c}\left[pc (\Sigma)^q+W\left(-pc(\Sigma)^{q}e^{-pc (\Sigma)^{q}}\right)\right]
.
\eea
Inserting this solution back into the equation for $\Sigma$ in Eqs.~(\ref{ps})  
we find
\begin{widetext}
\bea
\Sigma=G_1^q(\Sigma)\left[p +\sum_{q}\frac{q P(q)}{\Avg{q}}\frac{1}{c}(\Sigma)^{-q}W\left(-pc(\Sigma)^{q}e^{-pc (\Sigma)^{q}}\right)\right].
\label{op}
\eea

This equation can be written as $F(\Sigma,p)=0$. By imposing both $F(\Sigma,p)=0$ and $dF(\Sigma,p)/d\Sigma=0$ we can find the set of critical points $p=p_c$ 
in which discontinuities (jumps) of the function $\Sigma(p)$ take place. 
The  equation $dF(\Sigma,p)/d\Sigma=0$ reads as 
\bea
\frac{1}{p}
=
\sum_{q}\frac{q P(q)}{\Avg{q}}(1-e^{-c\sigma_q})\sum_q \frac{q(q-1)P(q)}{\avg{q}}\Sigma^{q-2}+cp G_1^q(\Sigma)\sum_{q<q_{max}} \frac{q^2P(q)}{\avg{q}}e^{-c\sigma_q}\frac{\Sigma^{q-1}(1-e^{-c\sigma_q})}{1-pc\Sigma^qe^{-c\sigma_q}}
,
\label{e220}
\eea
\end{widetext}
with $q_{max}=
[
-\log(pc)/\log\Sigma
]
$. Here $[...]$ stands for the integer part. 
Analysing Eq.~(\ref{op}) for the order parameter we can show that, in addition to a jump, the order parameter $\Sigma$ has a singularity at each $p=p_c,\Sigma=\Sigma_c$ where  both $F(\Sigma_c,p_c)=0$ and  $\left.dF(\Sigma,p)/d\Sigma\right|_{\Sigma=\Sigma_c,p=p_c}=0$.
For every topology of the network of network, 
we have 
\bea
\Sigma-\Sigma_c\propto (p-p_c)^{1/2},
\eea
similarly to 
multiplex networks \cite{Doro}. 

In the special situation in which the minimal superdegree $m$ of the local supernetworks is greater or equal to 2, i.e., $m\geq2$, each local supernetwork only has a giant connected component and does not have finite components.  
Therefore in this case we found that in the limit $M\to \infty$ the only viable solution for the order parameter is 
$\Sigma=0$ for any $p$ and $c$ 
and any superdegree distribution $P(q)$. In another special case in which $m=0$ there are layers that are not interacting  ($q=0$) with other layers. These layers can be treated separately without losing the generality of the treatment. Therefore the only non-trivial case is the case of $m=1$, which we explore in detail. 

\begin{figure*}[t]
\begin{center}
\centerline{\includegraphics[width=6.1in]{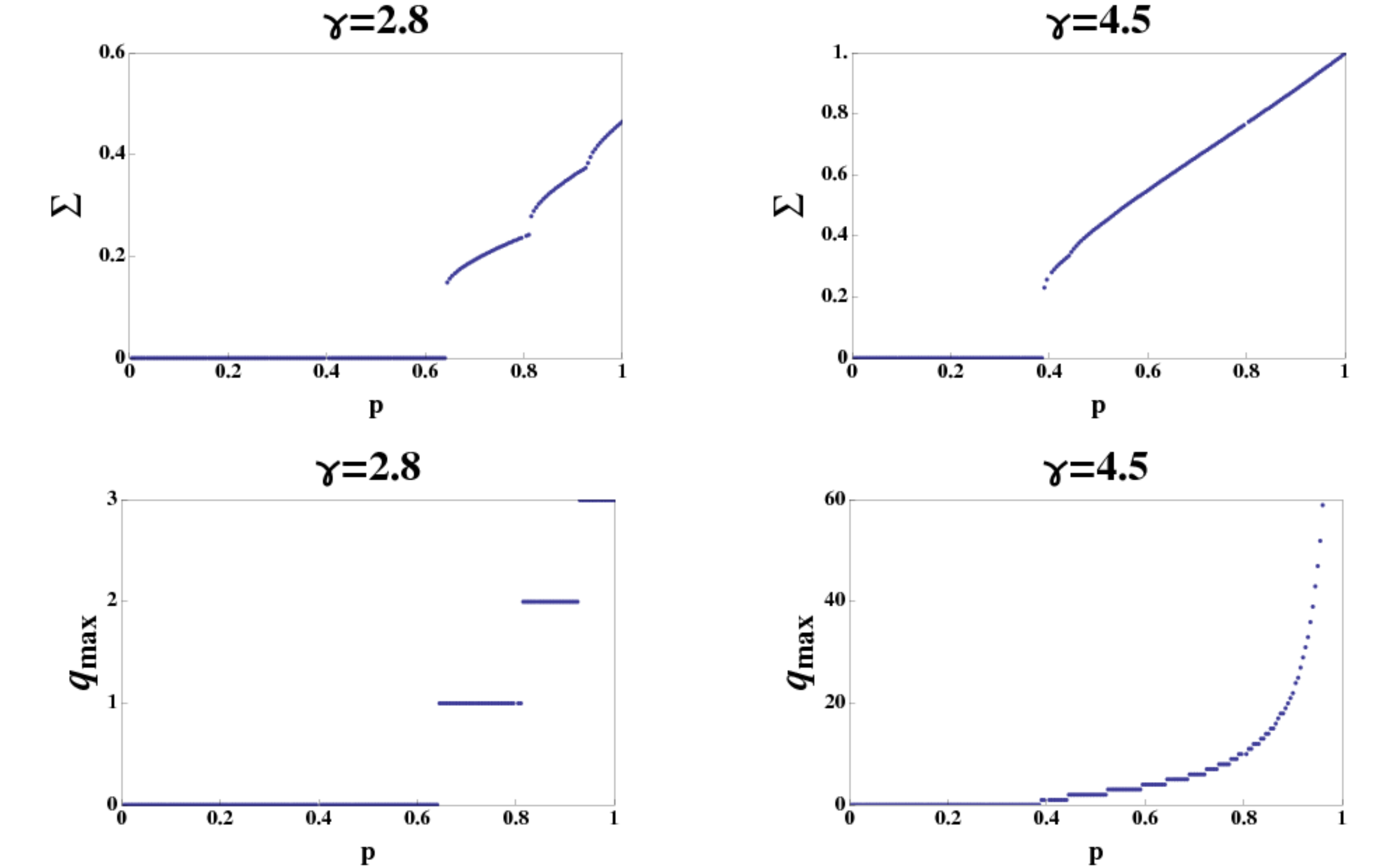}}
\end{center}
\caption{
Plot of the order parameter $\Sigma$ vs. $p$ and of the maximal superdegree of  percolating layers $q_{max}$ vs. $p$ for a configuration model with a Poisson $P(k)$ distribution  with average $\Avg{k}=c=20$ and a scale-free $P(q)$ distribution with $\gamma=2.8$ and $\gamma=4.5$, and with  minimal degree $m=1$ and maximal superdegree $Q=10^3$. 
These results are obtained by numerical solution of Eq.~(\ref{op}) for the order parameter. 
}
\label{fig:fig1a}
\end{figure*}
\begin{figure}[h!]
\begin{center}
\includegraphics[width=195pt]{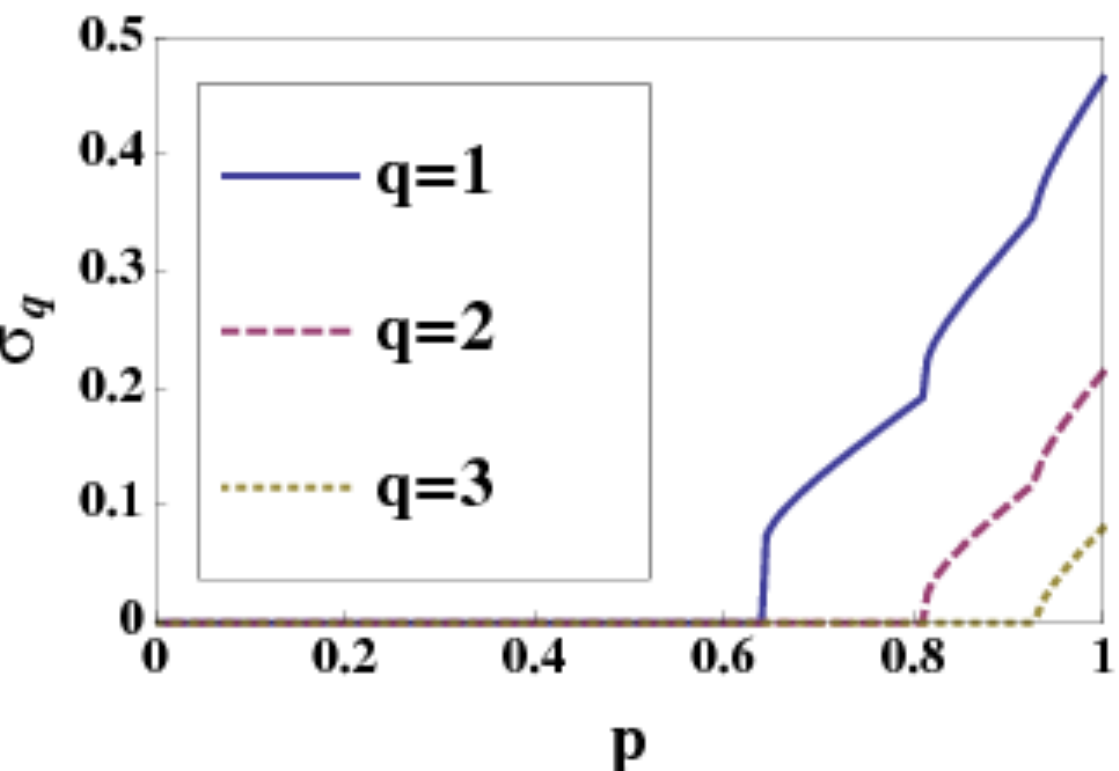}
\end{center}
\caption{
Plot of $\sigma_q=S_q$ (fraction of nodes in a layer of superdegree $q$, belonging to the mutual component) vs. $p$ for different values $q=1,2,3$ in the configuration model of the network of networks having a Poisson $P(k)$ distribution  with average $\Avg{k}=c=20$ and a scale-free $P(q)$ distribution with $\gamma=2.8$, minimal superdegree $m=1$ and maximal superdegree $Q=10^3$. For each value of $q=1,2,3$, $\sigma_q$ emerges discontinuously, with a jump, which becomes smaller and smaller with increasing $q$. The emergence of $\sigma_2$ is accompanied by a discontinuity of $\sigma_1(p)$. The emergence of $\sigma_3$ is accompanied by discontinuities of $\sigma_1(p)$ and  $\sigma_2(p)$. 
These curves are obtained by solving numerically Eq.~(\ref{op}) and substituting the result into Eq.~(\ref{e160}). 
}
\label{f2}
\end{figure}
\begin{figure*}[t]
\begin{center}
 \centerline{\includegraphics[width=4.7in]{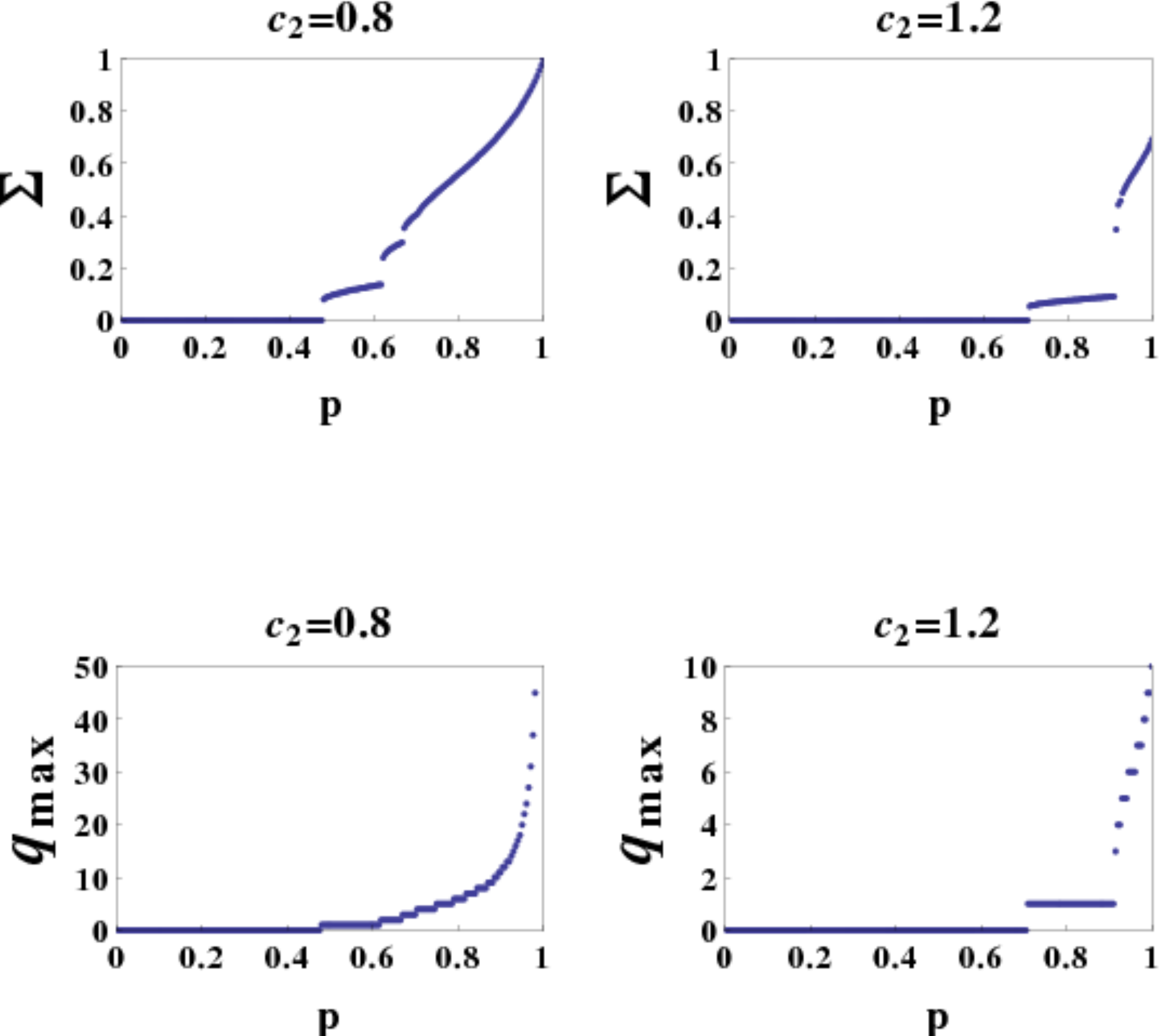}}
 \end{center}
  \caption{Plot of the order parameter $\Sigma$ vs. $p$ and of the maximal superdegree of  percolating layers $q_{max}$ vs. $p$ for a configuration model with a Poisson $P(k)$ distribution with $\avg{k}=c=50$ and a Poisson $P(q)$ distribution $c_2=0.8,1.2$ and $q\in[1,Q]$ with  maximal superdegree $Q=10^3$. 
The curves are obtained by numerical solution of Eq.~(\ref{op}). 
}
  \label{fig:fig2a}
\end{figure*}

We solve numerically Eq.~(\ref{op}) for the order parameter using various superdegree distributions $P(q)$ and a Poisson distribution $P(k)$ for connections within layers. This gives $\Sigma(p)$, from which we find the observable fraction $S_q(p)$ of the vertices in layers characterising by superdegree $q$ that belong to the  mutual component. Since the distribution $P(k)$ is Poissonian, $S_q(p)=\sigma_q(p)$ according to Eq.~(\ref{e160}).  In Fig.~\ref{fig:fig1a} we show the resulting order parameter $\Sigma$ vs. $p$ for a configuration model with a Poisson $P(k)$ distribution with $\avg{k}=c=20$ and a scale-free distribution $P(q)\simeq  q^{-\gamma} $ for $q\in[m,Q]$ with $\gamma=2.8$ and $\gamma=4.5$, minimal superdegree $m=1$ and maximal superdegree $Q=10^3$. The supernetwork with $\gamma=2.8$ is above the ordinary percolation phase transition in the supernetwork (i.e., it has a giant component) while the supernetwork with $\gamma=4.5$ is below the ordinary percolation phase transition (i.e., it consists only of finite components). 
Notice that if the supernetwork has no giant component, then $\Sigma(p=1)=1$, otherwise, $\Sigma(p=1)<1$.  
The order parameter displays a series of discontinuous jumps corresponding to the transitions in which layers 
with increasing values of $q$ start to percolate. 
In other words, in layers with higher and higher superdegree $q$, giant clusters of the mutually connected component emerge progressively. 
In order to see this, we observe that $\sigma_q=S_q=0$ for all values of $\Sigma$ such that $cp\Sigma^q<1$ (note that for power-law superdegree distributions 
the minimal degree is always $m \geq 1$). 
In Fig.~\ref{fig:fig1a} we plot the maximal superdegree of the percolating layers 
as $q_{max}=
[-\log(cp)/\log \Sigma]
$,  
where in this expression $[\ldots]$ indicates the integer part. From Fig.~\ref{fig:fig1a} it is clear 
that the discontinuities in the curve $\Sigma=\Sigma(p)$ correspond to the percolation transitions of layers of increasing superdegree $q_{max}$.
For $\gamma=2.8$ the first, second and third transitions occur at 
$(p_1,\Sigma_1)=(0.644285,0.138357)$, $(p_2,\Sigma_2)=(0.814055,0.266496)$, and $(p_3,\Sigma_3)=(0.927368,0.37778)$. 
We found these values by solving numerically Eqs.~(\ref{op}) and (\ref{e220}). 
One can see that for $\gamma=4.5$ 
the  activations of layers of increasing superdegree $q_{max}$ become much more rapid.
For these parameters the first,  transitions occur at 
$(p_1,\Sigma_1)=(0.390686,0.226456)$. 
For the case of $\gamma=2.8$, which we discussed above, we plot the observables $\sigma_q=S_q$ (fraction of nodes in a layer of superdegree $q$, belonging to the mutual component) vs. $p$ for different values of $q$, see Fig.~\ref{f2}. This figure demonstrates how giant clusters of the mutually connected component emerge progressively in layers with higher and higher $q$ as we increase the control parameter $p$. 
Note that, as is natural, each discontinuous emergence of a fraction of the mutual component nodes in layers of superdegree $q$ is accompanied by discontinuities of the dependencies $\sigma_{q'}(p)=S_{q'}(p)$ for all smaller superdegrees $q'<q$.

\begin{figure*}[t]
\begin{center}
 \centerline{\includegraphics[width=4.7in]{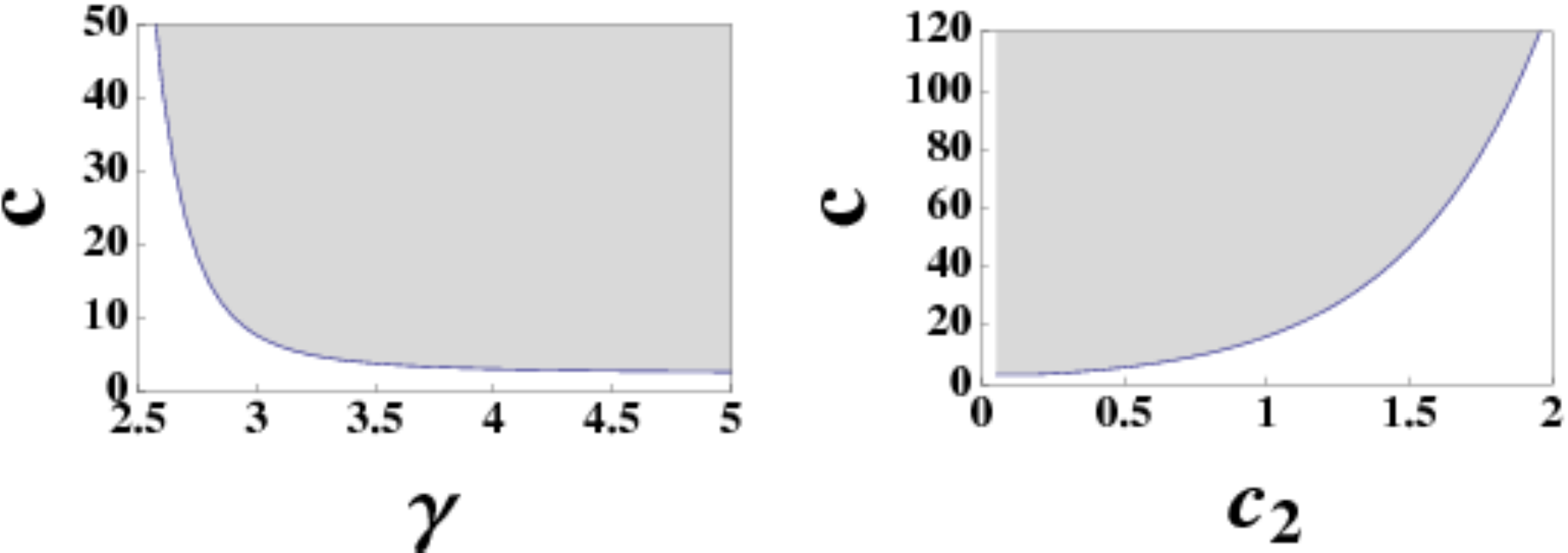}}
 \end{center}
  \caption{Phase diagram of a power-law supernetwork (left panel) and a Poissonian supernetwork (right panel) at $p=1$. The shaded region shows the phase with $\Sigma>0$, and in the white region, $\Sigma=0$. 
  The power-law super network has superdegree distribution $P(q)\propto q^{-\gamma}$ with $q\in[m,Q]$, the Poisson supernetwork has superdegree distribution $P(q)\propto c_2^q/q!$ with $q\in [m,Q]$. In both cases we have taken $m=1$ and $Q=10^3$. 
This phase diagram is obtained by numerical analysis of Eq.~(\ref{op}) for the order parameter.}
  \label{fig:fig3a}
\end{figure*}

As a second example of the  configuration model, we have taken a network of networks ensemble  with a Poisson $P(k)$ distribution characterized by  $\avg{k}=c$ and a Poisson distribution $P(q)=(c_2)^q e^{-c_2}/q!$, where  we take $q\in[1,Q]$, excluding the  layers with $q=0$ that are not interacting,    
and using the superdegree cutoff $Q=10^3$ for performing the numerical calculations. Solving numerically Eq.~(\ref{op}) we obtained the following results.  
In Fig.~\ref{fig:fig2a} we have chosen $c=50$ and $c_2=0.8, 1.2$  
as examples of the supernetwork below ($c_2=0.8$) and above ($c_2=1.2$) the ordinary percolation phase transition 
in the Poisson supernetwork.   
In order to see at which value of $p$ the layers with superdegree $q_{max}$ become percolating, in  Fig.~\ref{fig:fig2a} we also plot the maximal superdegree of the percolating layers $q_{max} = [-\log(cp)/\log \Sigma]$ vs. $p$. 
The transition points and the corresponding values of the order parameter are obtained by solving numerically Eqs.~(\ref{op}) and (\ref{e220}). 
For $c_2=0.8$ the first transitions are $(p_1,\Sigma_1)=(0.478781,  0.07489530)$, $(p_2,\Sigma_2)=(0.619669,0.232843)$, and $(p_3,\Sigma_3)=(0.667896,0.338005)$.
For $c_2=1.2$ the first transitions are $(p_1,\Sigma_1)=(0.714137,0.0502185)$, $(p_2,\Sigma_2)=(0.915646,0.204011)$, and $(p_3,\Sigma_3)=(0.918237,0.421259)$. 
For a range of the network of networks parameters, the giant mutual component is absent at any value of $p$, including $p=1$. 
In Fig.~\ref{fig:fig3a} we plot the phase diagram containing a phase in which $\Sigma=0$ (white region) and a phase in which $\Sigma>0$ (shaded region) for $p=1$. This phase diagram was obtained by numerical analysis of the equation for the order parameter. 
In particular, the phase diagram $(\gamma,c)$ is plotted for a 
scale-free supernetwork with power-law exponent $\gamma$ and superdegrees $q\in[m,Q]$ with $m=1$ and $Q=10^3$ and the phase diagram $(c_2,c)$ is plotted for a Poisson supernetwork with average superdegree $\avg{q}=c_2$ and the minimal and maximal superdegrees given respectively by  $m=1, Q=10^3$. As is natural, Figs.~\ref{fig:fig1a} and \ref{fig:fig2a} correspond to points within the shaded areas in Fig.~\ref{fig:fig3a}. 


\section{Percolation in the configuration model of Network of Networks with partial interdependence}

In this section we consider an interesting variation of 
the 
percolation problem in the configuration model of the network of networks.
In particular we assume that the superdegree adjacency matrix ${\bf a}$ has probability $P({\bf a})$ given by Eq.~(\ref{Pa}), while the probability of $P(\{s_{i\alpha}\})$ is given by Eq.~(\ref{Ps}).
In addition to that we assume that 
an interlink between replicas $(i,\alpha)$ and $(i,\beta)$ is damaged and permanently removed  with probability $1-r$.
In this case the nodes in a layer $\alpha$ with superdegree $q_{\alpha}=q$ can be interdependent on $n\in[0,q]$ other randomly chosen layers.
In this case the message passing equations are given by  
\bea
\hspace*{0mm}
\sigma_{i\alpha \to j\alpha}&=&
\!\!\!\prod_{\gamma\in {\cal C}({i,\alpha})\setminus \alpha}\!\left\{s_{i\gamma}\left[1-\prod_{\ell\in N_{\gamma}(i)}(1-\sigma_{\ell\gamma \to i\gamma})\right]\right\}
\nonumber 
\\[5pt]
&&\times s_{i\alpha}\left[1-\prod_{\ell\in N_{\alpha}(i)\setminus j}(1-\sigma_{\ell\alpha \to i\alpha})\right],\nonumber 
\eea
where ${\cal C}(i,\alpha)$ is the connected cluster of the local supernetwork of node $i$ when we consider only interdependencies, i.e., only the 
interlinks that are not damaged.
We can therefore indicate by $\sigma_q$ the average messages 
within a layer $\alpha$ of degree $q_{\alpha}=q$, obtaining 
\bea
\sigma_{q}&=&p\sum_{s}\sum_{n=0}^q\binom{q}{n}r^n(1-r)^{q-n}P(s|n)B^{s-1}
\nonumber 
\\[5pt]
&&
\times[1-G_1^k(1-\sigma_{q})].
\label{sqpr}
\eea
Here, $P(s|n)$ indicates the probability that a node $i$ in layer $\alpha$ with $q_{\alpha}=q$ and $n$ interdependent layers is in a connected component ${\cal C}(i,\alpha)$ of the local supernetwork of interdependent layers
of cardinality  $|{\cal C}(i,\alpha)|=s$. Moreover, $B$ is defined as
\bea
B=
p\sum_{q'}\frac{q' P(q')}{\Avg{q}}
[1-G_0^k(1-\sigma_{q'})]
.
\label{B2}
\eea
Similarly, the probability that a node $i$ in a layer $\alpha$ with superdegree $q_{\alpha}=q$ is in the mutually connected component $S_q=\Avg{S_{i,\alpha}}$ is given by 
\bea
S_q&=&p\sum_{s}\sum_{n=0}^q\binom{q}{n}r^n(1-r)^{q-n}
P(s|n)B^{s-1}
\nonumber 
\\[5pt]
&&\times[1-G_0^k(1-\sigma_{q})].
\label{Sqpr}
\eea
If the layers are not all formed by a single giant component or if $p<1$ we have  $B<1$, and therefore we can neglect in Eqs.~(\ref{sqpr})--(\ref{Sqpr}) the contribution coming from the giant component of the local supernetworks in the large $M$ limit. Therefore in Eq.~(\ref{sqpr}) we can 
replace $P(s|q)$ with the probability $P_f(s|q)$ that a node of degree $q$  belongs to a finite component of size $s$ in the supernetwork.  
Let us consider the quantity
$P(s)=\sum_{q}
[qP(q)/\avg{q}]
\sum_{n=0}^q\binom{q}{n}r^q (1-r)^{q-n}P_f(s|n)$. Since $P(s)$ only depends on the distribution of finite components we have 
\bea
P(s)&=&\sum_{q}\frac{qP(q)}{\Avg{q}}\sum_{n=0}^q\binom{q}{n}r^q (1-r)^{q-n}
\nonumber
\\[5pt]
&
\times
&
\!\!\!\!\!\!\sum_{s_1,s_2,\ldots s_n}\prod_{\ell=1}^nP(s_{\ell})\delta\left(\sum_{\ell=1}^qs_{\ell},s-1\right).
\eea
Therefore $P(s)$
has a generating function $H(z)=\sum_{s}P(s)z^s$ that satisfies the equation
\bea
H(z)=z G^q_1(rH(z)+1-r).
\label{Hr}
\eea
Moreover,
since 
\bea
P_f(s|n)=\sum_{s_1,s_2,\ldots s_n}\prod_{\ell=1}^nP(s_{\ell})\delta\left(\sum_{\ell=1}^ns_{\ell},s-1\right).
\eea
we find 
\bea
&&\sum_{s}\sum_{n=0}^q\binom{q}{n}r^q (1-r)^{1-n}P_f(s|n)B^{s-1}
\nonumber 
\\[5pt]
&&={[rH(B)+1-r]}^{q}.
\eea
Therefore Eqs.~(\ref{sqpr}), (\ref{Hr}), and (\ref{B2}) have the form   
\bea
\sigma_q&=&p\,\, [rH(B)+(1-r)]^q [1-G_1^k(1-\sigma_q)] 
,
\nonumber
\\[5pt]
H(z)&=&zG^q_1(rH(z)+1-r) 
,
\nonumber
\\[5pt]
B&=&
p\sum_{q'}\frac{q' P(q')}{\Avg{q}}
[1-G_0^k(1-\sigma_{q'})]
.
\eea
Putting $H(B)=\Sigma$ we find that Eqs.~(\ref{sq}) and Eqs.~(\ref{Sq}) become
\bea
\sigma_q&=&p\,\,(r\Sigma+1-r)^q[1-G_1^k(1-\sigma_q)] 
,
\nonumber
\\[5pt]
S_q&=&p\,\,(r\Sigma+1-r)^q[1-G_0^k(1-\sigma_q)] 
,
\nonumber
\\[5pt]
\Sigma &=&\left[p\sum_{q'}\frac{q' P(q')}{\Avg{q}}
[1-G_0^k(1-\sigma_{q'})]\right]
\nonumber 
\\[5pt]
&&\times \sum_{q'}\frac{q' P(q')}{\Avg{q}}(r\Sigma+1-r)^{q'-1}
.
\eea

Let us consider for simplicity the case in which each layer is formed by a Poisson network with average degree $\avg{k}=c$. Then the previous equations become
\bea
\sigma_q
&=&
S_q=p\,\,(r\Sigma+1-r)^q(1-e^{-c\sigma_q}) 
,
\nonumber
\\[5pt]
\Sigma&=&\left[p\sum_{q'}\frac{q' P(q')}{\Avg{q}}
[1-e^{-c\sigma_{q'}}]\right]
\nonumber
\\[5pt]
&&\times 
\sum_{q'}\frac{q' P(q')}{\Avg{q}}(r\Sigma+1-r)^{q'-1}
.
\label{psp}
\eea
In the case in which the local supernetwork is regular, i.e. $P(q)=\delta(q,m)$, we have $\Sigma[r\Sigma+1-r]\sigma_m=\sigma$  satisfying the following equation 
\bea
\!\!\!\sigma=p\left[\frac{1}{2}\left(1-r+\sqrt{(1-r)^2+4r\sigma}\right)\right]^{q}(1-e^{-c\sigma})
.
\eea
For an arbitrary distribution $P(q)$, the problem defined in Eqs.~(\ref{ps}) 
still has a single order parameter 
$\Sigma$. 
As in Section~\ref{s3}, 
the  first equation in Eqs.~(\ref{ps}) has solution expressed in terms of the principal value of the Lambert function $W(x)$, namely,  
\begin{widetext}
\bea
\sigma_q = \frac{1}{c}\left[pc (r\Sigma+1-r)^q 
+
W\left(-pc(r\Sigma+1-r)^{q}e^{-pc (r\Sigma+1-r)^{q}}\right)\right].
\eea
Inserting this solution back in the equation for $\Sigma$ in Eqs.~(\ref{ps})  
we find
\bea
\Sigma=G_1^q(r\Sigma+1-r)\left[p +\sum_{q}\frac{q P(q)}{\Avg{q}}\frac{1}{c}(r\Sigma+1-r)^{-q}W\left(-pc(r\Sigma+1-r)^{q}e^{-pc (r\Sigma+1-r)^{q}}\right)\right].
\label{op2}
\eea
Similarly to Section~\ref{s3}, we write this equation 
as $F_2(\Sigma,p,r)=0$. By imposing 
$F_2(\Sigma,p,r)=0$ and $dF_2(\Sigma,p,r)/d\Sigma=0$ we can find the set of critical points $p=p_c$ at which the function $\Sigma(p)$ turns out to be discontinuous. 
The  equation $dF_2(\Sigma,p,r)/d\Sigma=0$ reads as 
\bea
\hspace{-10pt}
\frac{1}{p}
=\sum_{q}\frac{q P(q)}{\Avg{q}}(1{-}e^{-c\sigma_q})\!\sum_q \frac{q(q{-}1)P(q)}{\avg{q}}[r\Sigma{+}1{-}r]^{q-2}
{+}cp G_1^q(r\Sigma{+}1{-}r)
\!\!\!\!\!\sum_{q<q_{max}}\!\!\!\! 
\frac{q^2P(q)}{\avg{q}}e^{-c\sigma_q}
\frac{[r\Sigma{+}1{-}r]^{q-1}(1{-}e^{-c\sigma_q})}{1-pc[r\Sigma{+}1{-}r]^qe^{-c\sigma_q}}.
\eea
\end{widetext}
where $q_{max}=[-\log(pc)/\log(r\Sigma+1-r)]$.
Note that in contrast to the case of $r=1$, for $r\neq 1$ all values of the minimal superdegree $m\geq0$ provide non-trivial network of networks. Indeed, for $r< 1$, even if $m\geq 2$, local supernetworks have finite connected components. 
In addition to the abrupt phase transitions, this  model displays also a set of  continuous phase transition for different $q$, where the order parameter $\sigma_q$ acquires a non-zero value. These transitions occur at 
\bea
p=p_c=\frac{1}{c(1-r+r\Sigma)^q}.
\eea
These transitions are only stable for $r$ below some special value $r_q$, and at $r=r_q$  
they become discontinuous.
The value $r_q$
can be obtained by solving simultaneously the following set of equations
\bea
\frac{1}{c(1-r+r\Sigma)^q}&=&p
,
\nonumber 
\\[5pt]
F_2(\Sigma,p,r)&=&0
,
\nonumber 
\\[5pt]
\frac{dF_2(\Sigma,p,r)}{d\Sigma}&=&0
.
\eea


\section{Conclusions}

In this paper we have characterized the robustness properties of the configuration model of network of networks by evaluating the size of the mutually connected component when a fraction $1-p$ of nodes have been damaged and removed.
The configuration model of network of networks is an ensemble of multilayer networks in which each layer is formed by a network of $N$ nodes, and where each node might depend only on its ``replica nodes'' on the other layers.
We assign to each layer $\alpha$ a superdegree $q_{\alpha}$ indicating the number of interdependent nodes of each individual node $(i,\alpha)$ of the layer $\alpha$ 
and take these $q_{\alpha}$ ``replica nodes'' in $q_{\alpha}$ uniformly randomly chosen layers, independently for each node $(i,\alpha)$. 
We have shown that  
percolation in
this ensemble of networks of networks demonstrate surprising features. 
Specifically, 
for low values of $p$, only the layers with low enough value of the superdegree are percolating, and as we raise the value of $p$ several discontinuous transition can occur in which layers of increasing value of 
superdegree $q$ begin 
percolate. 

Here we observe a sharp contrast to ordinary percolation in which nodes of high degree belong to the giant connected component (percolation cluster) with higher probability. This principal difference is explained by the definition of the mutual component according to which a node is in a mutual component only if all the nodes interdependent with this node belong to the mutual component. This condition makes more difficult the entrance into the mutual component for layers with a high superdegree. The non-trivial point here is that a layer of each given degree enters the mutual component not smoothly but through a discontinuous transition. 

The networks of networks which we considered in this paper differ from those we studied in our previous work~\cite{BD} in one key aspect. In the present work, interdependence links of different nodes of a layer are not lead to the same other layers as in Ref.~\cite{BD}, but they are distributed over the other layers essentially more randomly, independently for different nodes of a layer. We have found that this additional randomness dramatically change results and leads to new effects. We obtained our results assuming that the number $M$ of layers in the network is infinite. A more realistic case of finite $M$ is a challenging problem. We would also like to stress that multiple discontinuous phase transitions in models for complex networks is a rear but not unique phenomenon. For example, multiple discontinuous transitions were recently reported in another network model \cite{Wu:wjzcclh2014}.

One should note that interlinking of only ``replica nodes'' is actually a great, very convenient simplification which has enabled us to solve the problem analytically. Moreover, we have first assumed that all nodes in the same layer have equal superdegree (number of interdependencies), which is a strong constraint. On the next step, we however have removed this restriction by introducing the probability $1-r$ that an interdependence link is removed. 
We found that when a fraction $1-r>0$ of interdependent links are removed, each of these specific transitions can 
change from discontinuous to continuous.  

In summary, we have found novel percolation phenomena in a more general model of a network of networks than the network models which were considered previously. The multiple transitions, accompanying the expansion of mutual component over the layers of such a network of networks, are in dramatic contrast to ordinary percolation and to more simple interdependent and multiplex networks, e.g., for a pair of interdependent networks. 
We suggest that our findings should be valid even for more general networks of networks. 

{\it Note added in proof}
We recently considered a network of networks in which interlinks between each two layers connect randomly selected nodes and not only “replica nodes”.Remarkably, it turned out that the results for this model are exactly the same as in the present article.


\begin{acknowledgments}

This work was partially supported by the FCT
project PTDC/MAT/114515/2009 and the FET proactive IP project MULTIPLEX number 317532.

\end{acknowledgments}


\appendix

\section{Derivation of Eq. (\ref{mesf})}

In Ref.~\cite{BD} the percolation transition in a network of network in which all the local adjacency matrices are the same, i.e., $A^i_{\alpha\beta}=A_{\alpha\beta}~~~ \forall i$, was considered, implying  that the local supernetwork is the same for every node $i$.
In this setting, the percolation properties are determined by the message passing Eqs.~(\ref{mp1}) with  ${\cal N}_i(\alpha)={\cal N}_{\alpha}$ indicating the set of layers $\beta$ which are neighbors of layer $\alpha$ in any local supernetwork.
For this network of networks, it was shown in Ref.~\cite{BD} that Eqs.~(\ref{mp1}) can be written  as 
\\
 \bea
S'_{i\alpha \to i\beta} &=& \prod_{\xi\in {\cal C}({\alpha})}\left\{s_{i\xi}\left[1-\prod_{\ell\in N_{\xi}(i)}(1-\sigma_{\ell\xi \to i\xi})\right]\right\}
,
\nonumber 
\\[5pt]
\sigma_{i\alpha\to j\alpha}&=&
\!\!\!\prod_{\gamma\in {\cal C}({\alpha})\setminus \alpha}\!\left\{s_{i\gamma}\left[1-\prod_{\ell\in N_{\gamma}(i)}(1-\sigma_{\ell\gamma\to i\gamma})\right]\right\}
\nonumber 
\\[5pt]
&&\hspace*{-2mm}\times s_{i\alpha}\left[1-\prod_{\ell\in N_{\alpha}(i)\setminus j}(1-\sigma_{\ell\alpha\to i\alpha})\right]
,
\label{msfA}
\eea
where ${\cal C}(\alpha)$ indicates the connected component of the supernetwork to which layer $\alpha$ belongs.
Moreover in Ref.~\cite{BD} it has been  shown that  $S_{i\alpha}$ is given by 
\bea
S_{i\alpha}&=&s_{i\alpha}\prod_{\beta\in {\cal C}(\alpha)}\left[1-\prod_{\ell\in N_{\beta}(i)}(1-\sigma_{\ell\beta \to i\beta})\right].
\label{S2A}
\eea 
Now we observe that all the steps performed in Ref.~\cite{BD} to obtain Eqs.~(\ref{msfA}) and (\ref{S2A}) are in fact only operations acting in the local supernetwork of node $i$. It follows immediately that for the network of networks coming from the configuration models, the same equations should be valid, where we replace  ${\cal C}(\alpha)$ with  the connected component of the local supernetwork ${\cal C}(i,\alpha)$ of node $i$ passing through layer $\alpha$.
Therefore for the configuration model of a network of networks the solution to the message passing Eqs.~(\ref{mp1}) and (\ref{S}) is given by  Eqs.~(\ref{mesf}) and Eqs.~(\ref{S2}).




\end{document}